\documentclass[aps,prl,twocolumn,superscriptaddress]{revtex4}
\usepackage{natbib}
\usepackage{amsmath}
\usepackage{amsfonts}
\usepackage{graphicx}
\usepackage{dcolumn}
\usepackage{tabularx}
\usepackage{keyval}
\usepackage{multirow}

\begin{document}
\title{Scattering rate collapse driven by a van Hove singularity in the Dirac semi-metal PdTe$_{2}$}
\author{Erik van Heumen}\email{e.vanheumen@uva.nl}
\affiliation{van der Waals - Zeeman Institute, University of Amsterdam, Sciencepark 904, 1098 XH Amsterdam, the Netherlands}
\author{Maarten Berben}
\affiliation{van der Waals - Zeeman Institute, University of Amsterdam, Sciencepark 904, 1098 XH Amsterdam, the Netherlands}
\author{Linda Neubrand}
\affiliation{van der Waals - Zeeman Institute, University of Amsterdam, Sciencepark 904, 1098 XH Amsterdam, the Netherlands}
\author{Yingkai Huang}
\affiliation{van der Waals - Zeeman Institute, University of Amsterdam, Sciencepark 904, 1098 XH Amsterdam, the Netherlands}

\begin{abstract} 
We present optical measurements of the transition metal dichalcogenide PdTe$_{2}$. The reflectivity displays an unusual temperature and energy dependence in the far-infrared, which we show can only be explained by a collapse of the scattering rate at low temperature, resulting from the vicinity of a van Hove singularity near the Fermi energy. An analysis of the optical conductivity suggests that below 150 K a reduction in the available phase space for scattering takes place, resulting in long-lived quasiparticle excitations. We suggest that this reduction in phase space provides experimental evidence for a van Hove singularity close to the Fermi level. Our data furthermore indicates a very weak electron-phonon coupling. Combined this suggests that the superconducting transition temperature is set by the density of states associated with the van Hove singularity. 
\end{abstract}

\maketitle
\section{Introduction}
The investigation of topological phases of matter has reignited interest and resulted in new insights in the electronic behaviour of various relatively simple materials families. One class of materials that is of particular interest for applications, isdic the transition metal dichalcogenides (TMDs). These materials host various forms of topological matter often in combination with a form of spontaneous symmetry breaking. As such, the TMDs offer exciting opportunities to study topological phase transitions, something that was previously only really possible in low temperature quantum Hall systems. One such example is PdTe$_{2}$, which features both bulk Dirac points \cite{bahramy:2017, nohPRL:2017, clarkPRL:2018} as well as topological surface states \cite{liu:2015} in combination with a transition to a superconducting phase with T$_{c}$ = 1.4 - 2 K \cite{guggenheim:1961,fei:2017,wangSR:2016,lengPRB:2017}. The superconducting state is conventional with an isotropic superconducting gap \cite{lengPRB:2017, amitPRB:2018, das:2018, voermanPRB:2019}. One peculiarity observed in these experiments is that PdTe$_{2}$ supports a type-I superconducting state \cite{lengPRB:2017,salis:2018}, but recently some surface effects have been observed that leave open the possibility for an interesting surface superconducting state \cite{sirohiJPCM:2019, lengarXiv:2019}. Superconductivity in PdTe$_{2}$ is undisputedly of a conventional, electron-phonon coupling origin. Recently, it was pointed out that there is a van Hove singularity close to the Fermi level in PdTe$_{2}$ \cite{kimPRB:2018} and that this could have important consequences for the behaviour of the superconducting phase when PdTe$_{2}$ is intercalated \cite{hooda:2018} or put under pressure \cite{lengarXiv:2019}. 

In this paper we investigate the normal state properties of PdTe$_{2}$ using optical spectroscopic methods. The most prominent feature in the optical spectra is an optical gap that becomes visible in the low temperature, far-infrared reflectivity with an edge around 37 meV. From an analysis of the optical spectra we identify several features that are consistent with interband transitions between previously observed bulk type-II Dirac states. We use the extended Drude model to investigate the temperature dependent response of the low energy charge carriers and find that as temperature decreases, the optical scattering rate continuously decreases and collapses below our detection limit at temperatures under 60 K. We speculate that this arises from the complete occupation or depletion of a van Hove singularity that sits in close proximity to the Fermi energy. The same analysis demonstrates that the electron-phonon interaction in this material is likely very weak and thus we suggest that the superconducting transition is set by the large density of states associated with the van Hove singularity. 

\section{Methods}
Single crystals of PdTe2 have been grown using a modified Bridgman method \cite{lyons:1976} from high quality starting materials. Starting from 99.99 $\%$ pure Pd and 99.999 $\%$ pure Te, single crystals were grown in a quartz ampoule at 800 $^{\circ}$C. The superconducting characteristics of these crystals, as determined by various experiments, were previously reported in Ref.'s \cite{lengPRB:2017,salis:2018, voermanPRB:2019,lengarXiv:2019}. Extensive characterisation details are reported in Ref. \cite{lengPRB:2017} (see also the supplementary material Ref. \cite{lengPRB:2017}). Energy dispersive x-ray analysis using a scanning electron microscope demonstrates the expected stoichiometry and a homogeneous distribution of the Pd and Te content. Resistivity data shows a residual resistivity ratio, RRR, equal to 30 with a low value of the room temperature resistivity of 23 $\mu\Omega cm$. This points to the high quality of these crystals and the very low impurity content. The superconducting transition takes place at $T_{c}$ = 1.57 $K$. No signs of degradation have been reported for these crystals and oxidation is not expected to play a role. Crystals used in our study were cleaved minutes before sealing the UHV cryostat and, as explained next, all experiments have been repeated several times to confirm reproducibility. These measurements did not show any signs of degradation over the period of the experiments. Reflectivity data was measured in the energy range between 5 meV and 4.35 eV using a Vertex v80 Fourier transform spectrometer. The temperature dependence of the reflectivity was measured using a home build cryostat with a position stabilised sample holder \cite{tytarenkoSR:2015}. Spectra were collected while continuously cooling to base temperature (10 K) and warming up to room temperature. The rate at which we change temperature is 1.3 K/min and we record 1 spectrum every minute. Temperature cycles were repeated up to five times, depending on the frequency range, to ensure reproducibility and improve signal-to-noise ratios, in particular in the far-infrared and UV ranges of the spectrum. By combining these temperature cycles we obtain a dense sampling of the temperature dependent spectrum. We subsequently average all spectra within a 2 K interval. After the temperature cycles on the sample, $in-situ$ evaporation of a thin reference layer was used to determine the absolute reflectivity. The temperature cycles were subsequently repeated on this reference layer. The reflectivity spectra are thus obtained between 10 and 300 K in 2 K steps. The resulting reflectivity data was analysed using the RefFit software package developed by A.B. Kuzmenko \cite{kuzmenko:2001a,Kuzmenko:2005jh}. 

\section{Results}
\begin{figure}[htb]
\includegraphics[width=8.6cm]{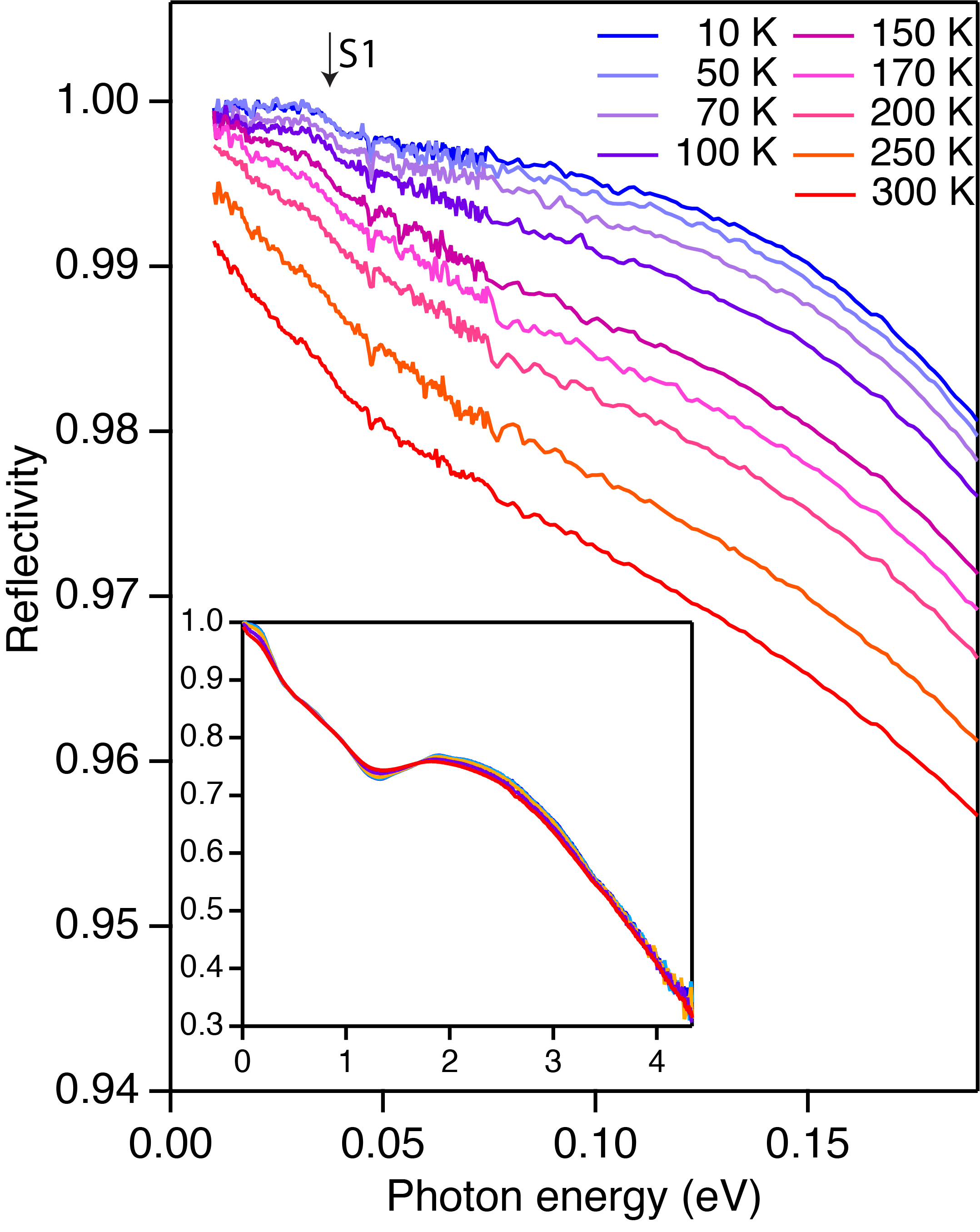}
\caption{(Color online) Reflectivity for selected temperatures. A step-like feature that develops at approximately 37 meV is indicated by S1, while a second feature 17 meV is indicated by S2. Both features become more prominent with decreasing temperature. The inset shows the reflectivity over the full range.}
\label{Refl}
\end{figure}
In Fig. \ref{Refl} the reflectivity for a few selected temperatures is shown. The far-infrared reflectivity of PdTe$_{2}$ displays characteristic signatures of a good metal: at room temperature the reflectivity follows a Hagen-Rubens behaviour from which we extract a DC resistivity of approximately 14 $\mu\Omega cm$, which is a little bit lower than what is observed in transport experiments \cite{lyons:1976, zhengPRB:2018, lengarXiv:2019}. As temperature decreases several features appear in the reflectivity. The most prominent of these is a step-like feature at 37 meV (labelled S1). This step gradually develops with decreasing temperature, forming a sharp plasma edge like structure around approximately 100 K. Remarkably, the reflectivity at the lowest temperatures is frequency independent below this edge and within our experimental uncertainty equal to 100 $\%$. It would be interesting to obtain information on the temperature dependence of the optical modes in PdTe$_{2}$ with an eye towards their importance for superconductivity. Neutron inelastic scattering experiments combined with calculations \cite{finlayson:1986} show that the zone center frequency for the E$_{1u}$ mode is 176 $cm^{-1}$ (21.9 meV). The A$_{u}$ mode observed by Finlayson \textit{et al.}, which is strongly coupled to the electrons and according to Kim \textit{et al.} \cite{kimPRB:2018} could be the main mode driving superconductivity is a bit lower in energy. Although there are weak features present in our data, we cannot positively identify them. If these are indeed optical modes, they are strongly screened by the free charge response. Above 0.2 eV the reflectivity decreases continuously (see inset), with some structure around 1.5 eV. The plasma edge falls outside our experimental window, but should be around 5 eV as we demonstrate below.
\\
\begin{figure}[htb]
\includegraphics[width=8.6cm]{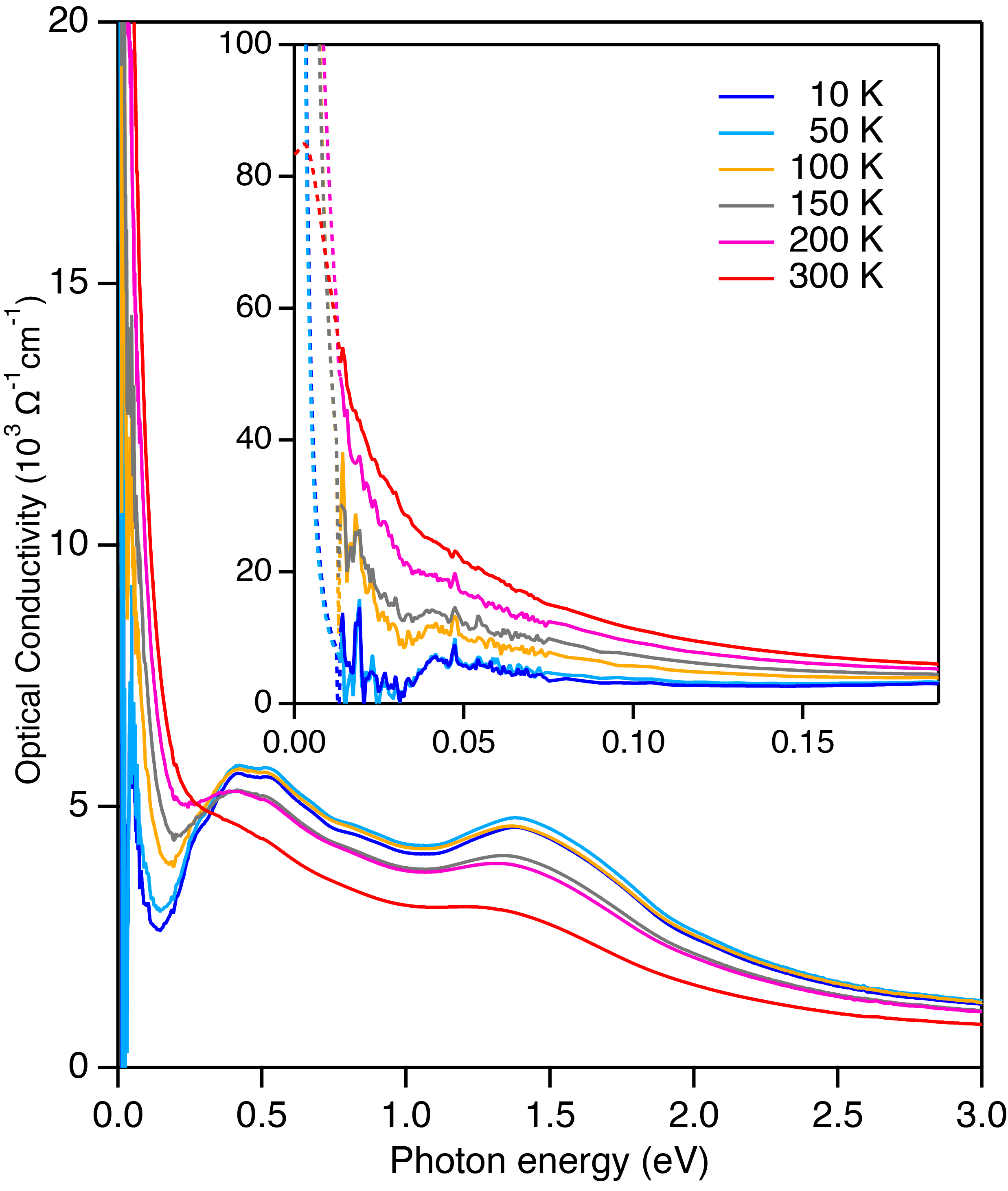}
\caption{(Color online) Optical conductivity at selected temperatures. We observe a midinfrared band centred around 0.5 eV. A second interband transition features prominently in the visible range (1.5 eV). The inset shows the low energy, optical response. Dashed lines are extrapolations based on the Drude-Lorentz model, while full lines are Kramers-Kronig transformed data. Note the formation of an optical gap around 0.03 eV at low temperature.}
\label{OptCond}
\end{figure}   
From the measured reflectivity one can obtain the optical conductivity using two methods: Kramers-Kronig transformation or through modelling. Although both methods should work equally well, modelling provides physically correct and flexible extrapolations outside the measured experimental range. We use a minimal Drude-Lorentz model with 6 oscillators at all temperatures. The model consists of a single Drude term, followed by a Lorentz oscillator with a central frequency that varies from 150 $cm^{-1}$ (at 10 K) to 400 $cm^{-1}$ (300 K) (i.e from 18 - 49 meV). Next we need two midinfrared oscillators, which are more or less temperature independent and have central frequencies of 3600 and 4300 $cm^{-1}$ (0.45 and 0.53 eV). The visible part of the spectrum can be modelled with a strong optical transition at 1.4 eV and a weaker one around 4 eV. The real part of the dielectric function extrapolates to a value $\varepsilon_{\infty}=3.5$.. 

Based on the Drude-Lorentz modelling we apply a variational dielectric function \cite{Kuzmenko:2005jh} that effectively carries out the Kramers-Kronig transformation. The resulting optical conductivity is displayed in Fig. \ref{OptCond}. As expected from the Drude-Lorentz model, we observe a midinfrared band centred around 0.5 eV and a strong transition around 1.5 eV. The midinfrared band has a significant fine-structure: we observe transitions at 0.27 eV, 0.42 eV, 0.52 eV, 0.65 eV and 0.83 eV. To determine the origin of these transitions, we turn to bandstructure calculations \cite{jan:1977, bahramy:2017, zhengPRB:2018, kimPRB:2018}. These indicate that the bands near the Fermi level consist predominantly of Te 5p orbitals. In particular, Ref. \cite{bahramy:2017} shows that along the $\Gamma$ - $A$ direction a band crossing takes place, resulting in a type-II bulk Dirac point at binding energy around -0.65 eV. At the same time surface states are observed with binding energies of -1.1 eV and -1.7 eV. To determine whether the optical transitions observed here arise from these bands, the optical selection rules need to be evaluated. Although the bands around the Fermi level are all derived from $p$-orbitals, the character of the hybridised bands is altered as a consequence of the strong crystal field and spin-orbit coupling in this material. The resulting bands carry different parities and consequently it becomes possible to have interband transitions between some of the resulting bands (for example, between the $R_{4}^{+}$ and the $R_{5,6}^{-}$ bands, following the notation of \cite{bahramy:2017}). It is therefore likely that the transitions observed here are optical transitions between the upper and lower branches of the bulk Dirac cone.

We next turn our attention to the temperature dependence of the free charge response, shown in the inset of Fig. \ref{OptCond}. As already hinted at by the very high reflectivity (Fig. \ref{Refl}), the optical conductivity displays a Drude peak that sharpens substantially with decreasing temperature. As the temperature decreases below 200 K, a dip in the Drude peak becomes visible at 32 meV, which eventually turns into the onset of an optical gap. This gap corresponds to the structure observed in the reflectivity labelled S1. The precise origin of the optical gap observed in our data is difficult to pinpoint with certainty, as there are multiple Fermi surface sheets and significant $k_{z}$ dispersion. We note that in a recent calculation \cite{kimPRB:2018} it was observed that there is a van Hove singularity (vHS) at an energy somewhat above the Fermi level. These calculations display a vHS approximately 0.12 eV above the Fermi level. However, the precise energy position of this vHS will depend on the electron density and details of the calculations. Indeed, we note that the same vHS appears in the calculations of Ref. \cite{bahramy:2017} right above the Fermi level. 

Presuming that the vHS in real materials remains somewhat above the Fermi level, it explains the temperature dependence of the free charge response and the opening of an effective optical gap. Starting from zero temperature and with the vHS approximately 30 meV above the Fermi level, we expect to observe a clean Drude response associated with the hole bands centred around the $\Gamma$-point. As temperature increases, thermal fluctuations will start to smear out the Fermi surfaces eventually resulting in partial occupation of the vHS states just above the Fermi level. The temperature where one expects this to start playing a role, is when 2-3$k_{B}T\approx E_{vHS}$. For the case at hand, this would imply that for temperatures between 120 K and 180 K a change in the temperature dependence due to the partial occupation of the vHS should become visible. Due to the large density of states associated with the vHS this should result in a strong enhancement of the scattering rate and potentially a reshuffling of the optical spectral weight. Note that this argument holds equally well if the vHS sits just below the Fermi level. In such a case, it is completely occupied at low temperature and a partial depletion of the vHS leads to a strong increase of the scattering rate.

\begin{figure}[htb]
\includegraphics[width=8.6cm]{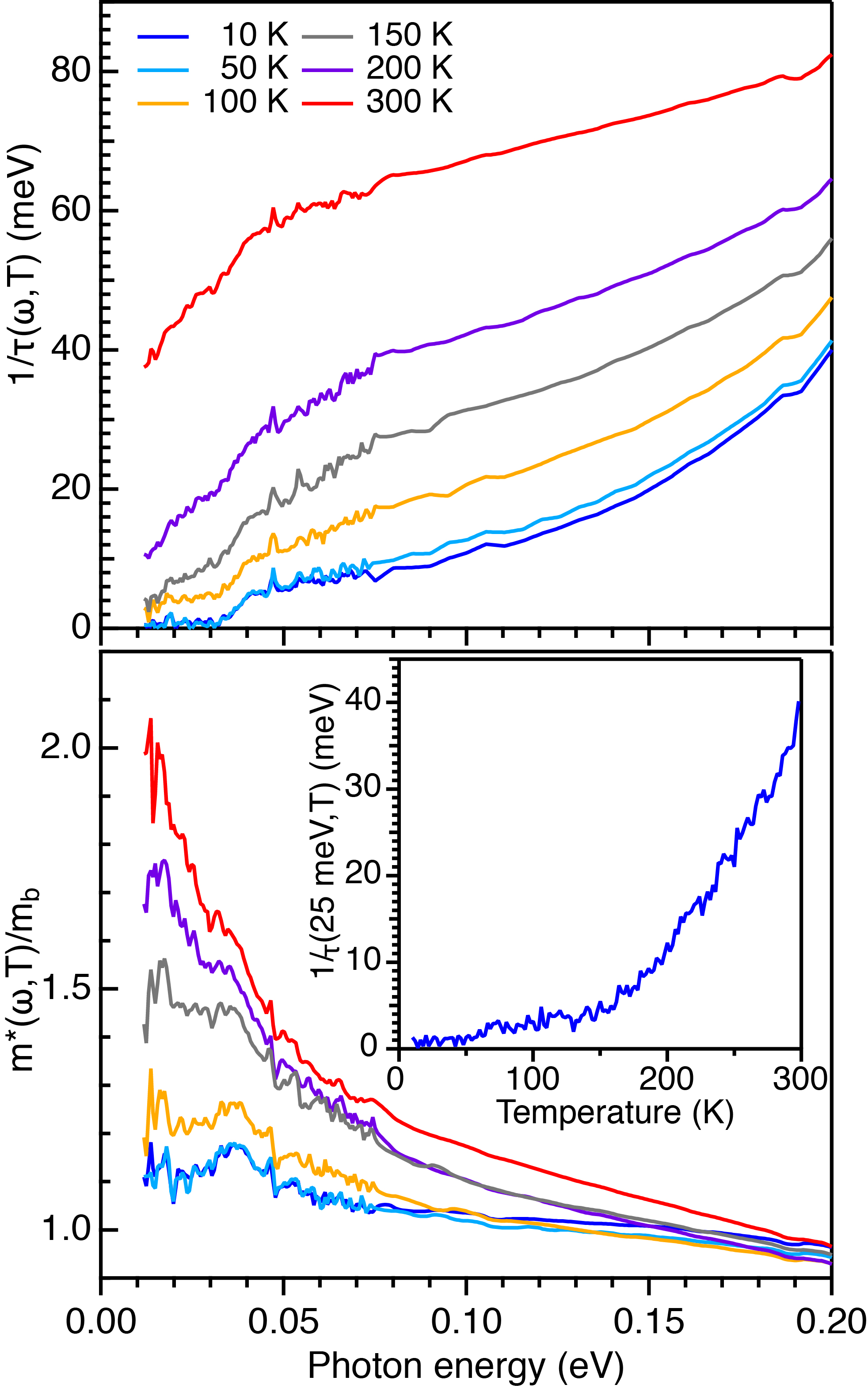}
\caption{(Color online) (a) Frequency and temperature dependent optical scattering rate for selected temperatures. The scattering rate displays a linear frequency dependence at high temperature, whereas it becomes frequency independent at low temperature and energy. (b) The mass enhancement factor show a strong decrease of the mass renormalisation with decreasing temperature. The inset displays the temperature dependent scattering rate at 0.025 eV.}
\label{tau}
\end{figure}

We investigate the enhancement of the scattering rate using the extended Drude model \cite{gotzePRB:1972}. Figure \ref{tau} displays both the real and imaginary part of the optical scattering rate, where the latter is commonly referred to as the mass enhancement factor. Fig. \ref{tau}a shows the frequency dependent optical scattering rate for a few selected temperatures. The inset of panel \ref{tau}b, displays the temperature dependence of the optical scattering rate at 25 meV. 
Starting at low temperature, we observe a frequency independent scattering rate (below 30 meV), almost below our detection limit, $1/\tau(0,T=10K)\approx$ 1$\pm$0.5 meV. Furthermore, the scattering rate is temperature independent up to 50 K as the inset in panel \ref{tau}b shows. This suggests that the scattering at low temperature and energy is completely dominated by impurity scattering. As temperature increases, the scattering rate acquires a frequency dependence that slowly turns linear over an extended energy range as clearly seen in the data above 200 K. The temperature dependence of the optical scattering rate (inset Fig. \ref{tau}b) shows a marked increase as temperature increases above 50 K. This increase in the scattering rate occurs when temperature becomes comparable to the energy of the lowest mode frequencies in the Eliashberg function $\alpha^{2}F(\omega)$ \cite{kimPRB:2018} and we therefore suggest that at this temperature electron-phonon scattering becomes relevant. At even higher temperatures ($\sim$ 150 K), a strong temperature dependence sets in and the scattering rate increases more or less linearly. The temperature where this takes place is indeed around the temperature where we expect the vHS to become partially occupied and the phase space for scattering to be significantly enhanced, thus confirming the first of the expected optical signatures of the presence of the vHS. The second feature, a redistribution of spectral weight is also observed in our data. This is clearly visible from figure \ref{OptCond} where at low temperature spectral weight is piling up below the energy scale set by the vHS. At low temperature the integrated spectral weight shows a plateau at low energies (40 meV) before further increasing as interband transitions start to contribute.  The integrated spectral weight in this plateau corresponds to a plasma frequency $\omega_{p}\approx$ 4.5 $\pm$ 0.1 eV. At room temperature the Drude response overlaps with interband transitions and we estimate that the spectral weight of the Drude peak decreases to a room temperature value of $\omega_{p}\approx$ 3.5 $\pm$ 0.1 eV. As the total spectral weight has to be conserved, this implies that new interband transitions become active as temperature is increased.  

The very low value of the optical scattering rate together with the plasma frequency determined above allows us to estimate the resistivity to be $\rho_{DC}(T=10 K)\approx$ 0.5$\pm$ 0.2 $\mu\Omega\cdot cm$, very close to what is reported  in recent transport experiments \cite{amitPRB:2018,lengarXiv:2019}. Combined with our observation of a frequency independent scattering rate up to the vHS energy scale $E_{vHS}$ suggests therefore that at low temperature the transport experiments are dominated by simple impurity and electron-electron scattering.

The observation of a frequency and temperature independent optical scattering rate at low temperature leave an interesting open question about the appearance of superconductivity below 1.5 $K$ \cite{guggenheim:1961, fei:2017, lengPRB:2017}. Since the mechanism of superconductivity in PdTe$_{2}$ is expected to be electron-phonon coupling driven, one would expect a signature of this interaction in spectroscopic experiments. In the superconducting state, such signatures are often visible as 'peak-dip-hump' structures, most famously observed in Giaever's first tunnelling experiments \cite{giaever:1962}. Similar features have also been observed in the far-infrared spectra of superconductors \cite{joyce:1970} and these are often taken as important clues to the mechanism underlying the superconducting state. Our observations of a featureless and temperature independent scattering rate suggest that the electron-phonon coupling is very weak below 50 $K$. Using the standard BCS result for T$_{c}$ = 1.14$\hbar\omega_{D} e^{(1/N(0)V)}$, our experimental data indeed imply an important role for the vHS in driving the superconducting instability as suggested in Ref. \cite{kimPRB:2018}. The highest phonon frequency to be expected in PdTe2 is of order 22 meV \cite{kimPRB:2018}. One would expect some change in the frequency dependence of the scattering rate around this energy. Even if the dominant electron-phonon coupling would be below our experimental energy range, we should detect a corresponding frequency dependence of the optical scattering rate in our experimental window. A final important indication for the absence of significant electron-phonon interaction is given by the mass enhancement factor, displayed in Fig. \ref{tau}b. There appears to be some mass enhancement at room temperature, but as the temperature decreases the mass enhancement decreases as well and reaches a minimum of approximately $m^{*}/m\approx$ 1.1 at $T=$ 10 $K$. 

It is interesting to consider the role of the vHS in other dichalcogenide systems, such as PtTe$_{2}$ and PtSe$_{2}$. Bandstructure calculations have been performed for PtTe$_{2}$ in Refs. \cite{zhengPRB:2018} and  \cite{kimPRB:2018}. Both calculations indeed show that the vHS appears at higher energy, e.g. further away from the Fermi level, which would result in a reduction of N(0).  If there is an important role for the vHS in determining T$_{c}$, PtTe$_{2}$ should have a lower T$_{c}$ and is indeed not superconducting. Non- superconducting PtSe$_{2}$ has been studied in ref. \cite{chengPRM:2017}, where a vHS appears around 0.5 eV above the Fermi level. These authors find that upon electron doping, their calculations predict a similar T$_{c}$ as for PdTe$_{2}$ (T$_{c}$ = 2.15 K). The authors indicate that the superconducting transition is driven by a softening of the acoustic phonon branch, but we note that for the case where the doping is changed by 0.5 electron/unit cell, the vHS comes precipitously close to the Fermi level.  

To summarise, we have presented optical spectra of the superconductor PdTe$_{2}$. The optical data features several midinfrared and visible range optical transitions, some of which may be related to optical transitions within the bulk Dirac cone. In the far-infrared we observe a sharp plasma edge like feature at low temperature temperature. A strong reduction of the optical scattering rate and a reshuffling of spectral weight provide evidence that this edge is the optical signature of a van Hove singularity some 30 meV above the Fermi level. This observation combined with a frequency independent optical scattering rate provides experimental support for a superconducting transition driven by a large density of states associated with a van Hove singularity.

\section{Acknowledgements}
We would like to thank A. de Visser for useful comments during the prepartation of this paper. This work is part of the research programme Strange Metals (grant number 16METL01) of the former Foundation for Fundamental Research on Matter (FOM), which is financially supported by the Netherlands Organisation for Scientific Research (NWO).

\bibliography{paper_library}
\end{document}